\documentclass[12pt,preprint]{aastex}

\newcommand\ta{\tau}

\newcommand\ph{\Phi}

\newcommand\bna{\bold{\nabla}}

\newcommand\<{\langle}
\renewcommand\>{\rangle}

\newcommand\beq{\begin{equation}}
\newcommand\eeq{\end{equation}}
\newcommand\bea{\begin{eqnarray}}
\newcommand\eea{\end{eqnarray}}
\newcommand\bal{\begin{align}}
\newcommand\eal{\end{align}}

\newcommand\fr{\frac}

\newcommand\cd{\cdot}

\newcommand\bk{\bold{k}}

\newcommand\br{\bold{r}}

\newcommand\bzero{\bold{0}}

\renewcommand\bal{\bold{\alpha}}

\newcommand\bth{\bold{\theta}}

\begin{document}

\title{Strong lensing time delay: a new way of measuring
cosmic shear}

\author{Richard Lieu}
\affil{Department of Physics, University of Alabama, Huntsville, AL 35899.}

\begin{abstract}

The phenomenon of cosmic shear, or distortion of images of distant sources
unaccompanied by magnification, is an effective way of probing the
content and state of the foreground Universe, because light rays do
not have to pass through matter clumps in order to be sheared.
It is shown that the delay in the arrival times between two
simultaneously emitted photons that appear to be arriving from a pair of
images of a strongly lensed cosmological
source contains not only information
about the Hubble constant, but also the long range gravitational
effect of galactic scale mass
clumps located away from the light paths in question.  This is therefore
also a method of detecting shear.
Data on time delays among a sample of
strongly lensed sources
can provide
crucial information about whether extra dynamics beyond gravity and
dark energy are responsible for the global flatness of space.
If the standard $\Lambda CDM$
model is correct, there should be a large dispersion
in the value of $H_0$ as inferred from the delay data by
(the usual procedure of) ignoring
the effect of all other mass clumps except the strong lens itself.
The fact that there has not been any report of a
significant deviation from the $h =$ 0.7 mark during any of the $H_0$
determinations by this technique may already be
pointing to the absence of the random effect discussed here.
\end{abstract}

\section{Introduction; time delay anisotropy from primordial matter distribution}

$\Lambda$CDM cosmology models the near Universe in terms of (a) the
gravity of {\it embedded} mass clumps in (b) a smooth `cosmic
substratum' of expanding space. While inflation may ensure an
Euclidean mean geometry,  fluctuations caused by the (a) phenomenon
operating over Hubble scales ought to be observable. It would be
very important, therefore, to test if the statistical effect of the
gravity of virialized structures distributed throughout the near
Universe exists. Such an effect can manifest itself as a random
delay in the arrival times of two light signals emitted
simultaneously at separate positions and detected by the same
observer O, after having propagated through different paths.

A calculation of the (finite) variance in such a delay was provided
by Lieu \& Mittaz (2007), who assumed a smooth Universe perturbed by
primordial matter fluctuations of power spectrum $P(k)$. Here we
simply sketch the essential steps on how it is done.  Denote the
gravitational perturbation of an otherwise zero curvature
Friedmann-Robertson-Walker Universe (as inferred from WMAP1 and
WMAP3, viz.  Bennett et all 2003 and Spergel et al 2007) by $\Phi
(x, {\bf y})$, where the $x$-axis is aligned with the light path and
${\bf y}$ is a vector along some direction transverse to ${\bf x}$.
If the angle one ray makes w.r.t. the other at O is $\theta$ and the
comoving light pathlength is $D$, the relative delay in the arrival
conformal time may be written as
 \beq \ta(\bth)-\ta(\bzero)= \frac{1}{c^3}
 \int_0^D\,2x'\bth\cd\bna\ph(x',\bzero)dx' +  \frac{1}{c^3}
 \int_0^D\,{x'}^2 (\bth\cd\bna)^2
 \ph(x',\bzero)dx' + \cdots, \eeq
where $\bna$ is the gradient operator transverse
to the vector ${\bf x}$, viz. along the ${\bf y}$ direction.

The variance in the difference between the delays in the two signals
has its lowest order term ensuing from the first integral in
Eq. (1), as
\beq
[\delta\tau (\bth)]^2 = \frac{\bth^2}{c^6} \int_0^D 2 x' dx' \int_0^D 2x'' dx''
\hat\theta_i \hat\theta_j \<\nabla'_i \Phi(x',\bzero)
\nabla''_j \Phi(x'',\bzero)\>,
\eeq
where $\<\nabla'_i \Phi \nabla''_j \Phi\>$ is the correlation function
between the two spatial gradients of $\Phi$, with the indices $i,j$
denoting the two transverse
directions $(0,1,0)$ and $(0,0,1)$, and summation over
repeated indices is implied.  The important point about
Eq. (2) is its dependence on $\bth^2$, leading
to a standard deviation $\delta\tau(\bth) \sim \theta$.  This zeroth
order contribution to $\delta\tau$, though large, has no  observational
consequence because it depicts a {\it coherent delay}
$\delta\ta(2\bth) = 2\delta\ta(\bth)$
which, as will be explained in the next section (see also
Bar-kana 1996 and Seljak 1994) simply causes
a  global absolute shift in the angular positions of images without
changing relative positions.
The next order contribution to $\delta\tau(\bth)$ would come from the
second integral in Eq. (1), i.e. $\delta\tau(\bth) \sim \bth^2$.  It
depicts the genuinely random excursion in
the {\it relative} delay between the two rays which
in principle is observable.

\section{The impossibility of observing the coherent time delay between the light curves of strong lensing multiple images}

The best way of demonstrating this impossibility is
by means of a concrete example.  Consider the two-dimensional
problem of Figure 1a, where all light rays
are confined to the $xy$-plane.  Let a
spherically symmetric lensing system be at
comoving position $(d,0,0)$,
with the observer at the origin, and let a source S at distance
$D=2d$ cause two images to appear on opposite sides of S, say at angular
positions $\theta_+$ and $-\theta_-$ (S is slightly off the
$x$-axis).  To begin with, suppose let were no gravitational
perturbations anywhere
near the lines of sight.  Then the distances of closest approach are
$b_\pm\approx d\theta_\pm$, assuming that $GM\ll b_{\pm}$.  Moreover, by
the symmetry of the problem, the rays beyond the lens meet the line from
lens to source at the same angles $\pm~\theta_\pm$.  The angles of
deflection of
the rays are $4GM/b_{\pm}$.  We shall not need the actual formulae for
position of the source, and the time delay between the two signals.

Next, in Figure 1b we introduce another mass affecting the
`observer's half' of the light
paths, in the plane of both paths, but to one side, bending
the rays in the same
direction.  Assume that the mass is not \emph{too} close, so that its
gravitational field may
be described by a linearly varying potential, of the form
$\Phi=-\phi(x)y$, where $\phi(x)$
is a smooth function peaked in some region of
$x$.  Then the null geodesic equation reduces to
$$\frac{d^2y}{dx^2}=-\frac{2}{c^2}\frac{\partial\Phi}{\partial y}=
\frac{2}{c^2} \phi(x),$$
with solutions
$$ y_\pm=\pm~ x\theta_\pm+ \frac{1}{c^2} \int_0^x 2(x-x')\phi(x')dx'. $$
The slopes of the two curves when they reach the lens are given by
 $$ y'_\pm=\pm~\theta_\pm+2\int_0^d \phi(x')dx', $$
i.e. the angle between them remains at the previous value of $\theta_+ +
\theta_-$.
We may picture the
wave fronts as moving backwards from the observer whilst always maintaining
orthogonality with the direction of propagation.
The rays are bent upwards, and the wavefronts in the upper ray have less speed
than those in the lower one (since the latter experiences
weaker potential), by just the
amount needed to ensure this condition is satisfied.

Given that what we see in Figure 1b is
the same as that in Figure 1a, the lensing mass L
must also be in a slightly different position, moved upwards
along the lensing plane by the
corresponding amount, $$ \delta y_{{\rm L}} =
\frac{1}{c^2} \int_0^d 2(d-x')\phi (x')dx'$$.
Unless there are further masses
affecting the propagation on the far side of the lens (i.e. the parts of
the light paths between the lens and the source)
the remainder of
the diagram is exactly as before, except for being rotated by the small
angle, $$ \psi =
\frac{2}{c^2} \int_0^d \phi (x')dx'$$ as shown in Figure 1b.  Specifically
if the perturbing mass is displaced in the $+y$ (or $+z$)
direction, the rotation will be about an axis parallel to $z$ (or $y$),
and in the sense of $+y$ (or $+z$).

On the far side, we could similarly trace
wavefronts of the signal propagating from the source.  We can think of the
time delay difference as occurring close to L, between
wavefronts of rays propagating from both ends.  Owing to the slight
misalignment $\alpha$
between the source and lens, this time delay difference will not
be zero, but it will be exactly the same for Figure 1b and Figure 1a.
There is no extra contribution to the difference
from the perturbing mass, in
the context of our lowest order (linear, or coherent) theory.

%\clearpage
\clearpage
\begin{figure}[H]
\vspace{-3cm}
\begin{center}
\hspace{-2cm}
\includegraphics[angle=0,width=7in]{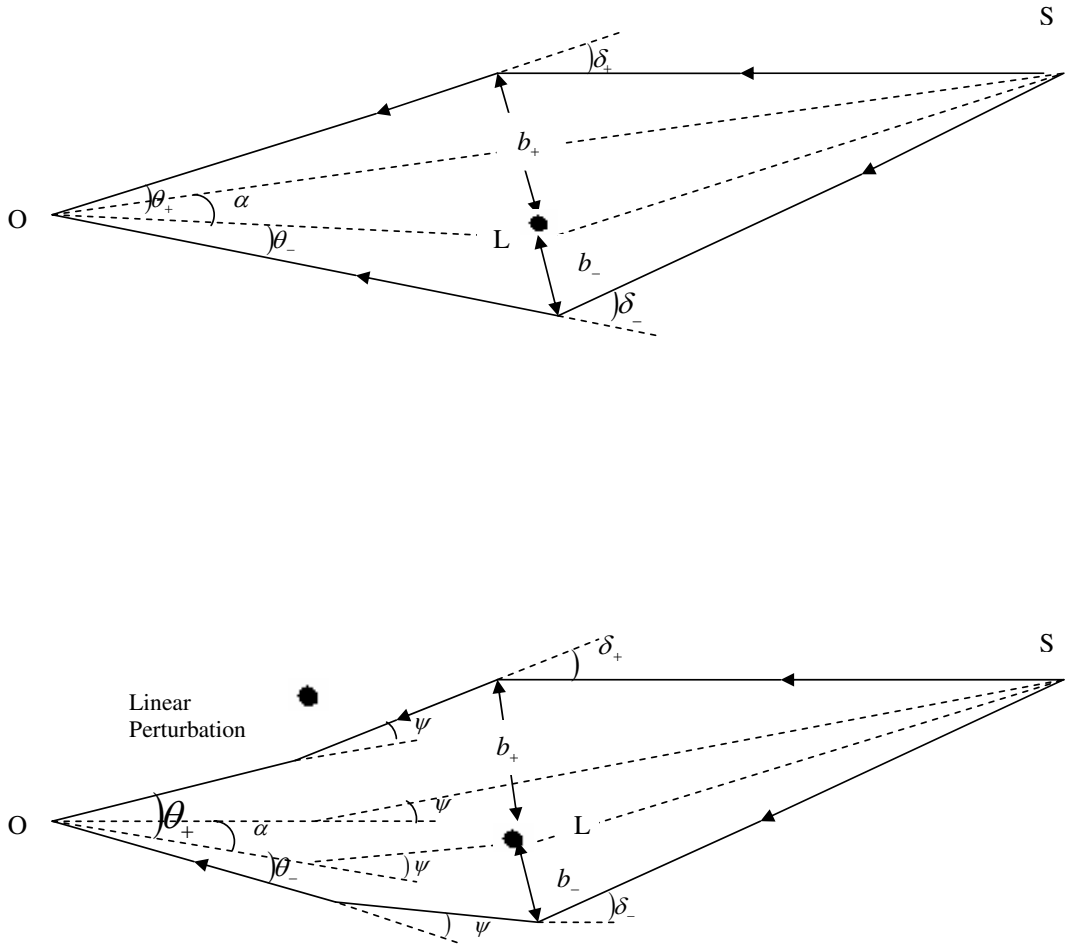}
\vspace{-8cm}
\end{center}
\caption{The propagation of light signals from a source to an
observer, with a strong gravitational lens in between.
The top sketch depicts the situation where the lens is the
only disturbance, while in the bottom sketch the effect of
a remotely located second mass is also present on the plane of
the two rays.  The dotted lines represent virtual rays going
directly from source and lens to the observer, with the bottom
sketch showing how these rays are deflected by the linear perturbation.}
\end{figure}
\clearpage

\section{Time delay anisotropy from a statistical ensemble of galaxies}

The analysis in section 1 of light propagation time through a
Universe of mass fluctuation could be continued with a calculation
of the correlation function between potential gradients, e.g. Eq.
(2) for the lowest order form of such a function, by expressing the
integral in terms of the matter power spectrum $P(k)$, \beq
\<\nabla'_i\ph(\br')\nabla''_j\ph(\br'')\> =
 \fr{9\Omega_m^2 H_0^4}{32\pi^3}
 \int \fr{d^3\bk}{k^3}k_ik_je^{i\bk\cd\br} P(k)
\eeq
where $\br = \br' - \br''$ and
\beq
P(k) = \frac{8\pi^2}{9\Omega_m^2 H_0^4}\frac{d}{d\ln k}(\delta\Phi_k)^2,
\eeq
with $\delta\Phi_k$ being the standard deviation of the potential
over length scales $2\pi/k$.

Now the objective of this paper is to investigate whether time delay
measurements can probe the mass distribution in the near ($z
\lesssim 1$) Universe where non-linear matter clumping is important,
because similar considerations in the context of primordial matter
that fills the high redshift Universe have already been made, with
the conclusion that such forms of matter causes negligible
additional delays (Surpi et al 1996,  Bar-kana 1996, Seljak 1994).
One could persist with the approach of Eq. (1) by adopting a
modified primordial matter spectrum $P(k)$ that includes an
`extension' to the non-linear, or large $k$, regime, (Peacock \&
Dodds 1994, Smith et al 2003), except that concerning the effect of
widely spaced and compact clumps on two closely separated light rays
it actually makes more sense to calculate directly the time delay
induced by the gravitational perturbation of a random ensemble of
clumps. This is because of several reasons: (a) while the
correlation function of Eq. (3) is relatively simple for the
unobservable lowest order term of $\delta\tau$ (i.e. sections 1 and
2) the observable next order term has a much more complex form; (b)
the shape of $P(k)$ is unreliable on sub-Mpc scales;  (c) the use of
a Poisson clump distribution is justified by the `nearest neighbour'
interaction phenomenon.  More elaborately the differential time
delay between two closely spaced rays is due mostly to proximity
galaxies located at distances sufficiently small from the two rays
in question for mass clustering (or compensation) to exert any
significant modification - see below.

Now even if  mass correlation over larger scales can be neglected by
appealing to Poisson clumps, there still is a form of $P(k)$ to
represent this kind of matter inhomogeneity (Eq. 23 of Peebles
1974). Owing to reason (a) above, however, it is much easier to
start from first principles. Let us return to the one-way Shapiro
time delay when light skirts a mass clump $m$ at `impact parameter'
$y$, or more precisely transverse position ${\bf y}$ (with
components along the conventional $(0,1,0)$ and $(0,0,1)$
directions, both perpendicular to ${\bf i} = {\bf \hat{x}}$) w.r.t.
the unperturbed ray, which is
\begin{equation}
\tau = \frac{2Gm}{c^3} \ln\left(\frac{4 x_s x_{ls}}{{\bf
y}^2}\right),
\end{equation}
where $x_s$ and $x_{ls}$ are the observer-source and  clump-source
distance respectively. Take a pair of rays with separation $\delta y
\ll y$, the difference in delay between them is
%\begin{align}
\begin{eqnarray}
\delta\tau & = & -\frac{2Gm}{c^3} \ln\left[\fr{{\bf y}^2+2{\bf
y}\cd\delta {\bf y}
+(\delta {\bf y})^2}{{\bf y}^2}\right] \nonumber \\
&  =  & -\frac{2Gm}{c^3} \left[\fr{2{\bf y}\cd\delta {\bf y}}{{\bf
y}^2}+ \fr{(\delta {\bf y})^2}{{\bf y}^2} -2\left(\fr{{\bf
y}\cd\delta {\bf y}}{{\bf y}^2}\right)^2 -2\fr{{\bf y}\cd\delta {\bf
y}}{{\bf y}^2}\fr{(\delta {\bf y})^2}{{\bf y}^2}
+\fr{8}{3}\left(\fr{{\bf y}\cd\delta {\bf y}}{{\bf y}^2}
\right)^3+\dots\right]
\end{eqnarray}
%\end{align}
If geometry is globally flat, the comoving separation $\delta y$
between the two rays at any position of comoving distance $x$ from
the observer O is given by
\begin{equation}
\delta y = x\theta = \frac{xd}{x_s},
\end{equation}
where $\theta$ is the angle subtended at O by two point sources of
distance $d$ apart, both being at the same comoving length $x_s$
away from O, to which our two rays map back.  These two points could
even mark the positions of a pair of strong lensing images, when the
light rays associated with the images are perturbed by external mass
clumps, in which case $x_s=D_l$, the distance to the lensing plane,
and $\delta\tau$ is the relative delay in the light arrival times
between the two images.

The lowest order observable effect is the incoherent contribution to
the relative delay between the two rays from each clump, which
originates from the {\it second} spatial derivative of the clump's
gravitational potential. From section 2 and Eq. (1) we that this is
the contribution arising from the $\delta\tau \sim \theta^2 \sim
(\delta y)^2$ terms of Eq. (6). In this light, it is clear that only
the 2nd and 3rd terms on the right side of Eq. (6) are relevant.
Thus, when we square the equation to form the variance, we obtain
\begin{equation}
(\delta\tau_{{\rm incoherent}})^2 = 2\left(\frac{Gm}{c^3}\right)^2
\left[\frac{(\delta {\bf y})^2}{{\bf y}^2}\right]^2 + {\rm
higher~order~terms},
\end{equation}
where the angle averages employed to go from Eq. (6) to Eq. (8) were
$\<\cos^2\vartheta\> =$ 1/2 and $\<\cos^4\vartheta\> =$ 3/8, with
$\vartheta$ being the angle between ${\bf y}$ and $\delta {\bf y}$.

For the rest of this section we shall indeed focus our attention
upon one manifestation of shear: the delay in photon arrival times
between two strong lensing images.   Under this scenario a random
walk arises as a result of the two rays skirting all the other
clumps on either side of the light path.  Their loci then become
like two long snakes (Hamana et al 2005, Gunn 1967a,b) as the rays
are deflected, largely in tandem, though there is {\it always} a
small and random relative change in directions, which is the {\it
same} shear phenomenon as the incoherent relative delay caused by
each clump.  Both these relative processes of deflection and delay
accumulate like $\sqrt{N}$ from one clump to the next if the clump
distribution is Poisson, as assumed.

Thus, specifically concerning time delay, as the rays continue their
journey the variances $(\delta\tau)_{{\rm incoherent}}^2$ from all
the clump encounters add, so that to evaluate the total excursion in
the arrival time difference one must perform a cylindrical
integration with the axis along the $x$-direction.  There is however
one subtlety here.  Since $y$ is the comoving impact parameter its
value for deflections at finite $z$ was smaller by by the factor
$1+z$.  Fortunately this factor cancels out when all the components
of the integrand are assembled, though the same does not happen when
one computes image distortions by shear, as we shall see.  The
cumulative variance for a Poisson ensemble of clumps of comoving
number density $n$ (i.e. neglecting {\it with caveats} the evolution
of clump properties, see below) is obtained by an integration down
the light path to be
 \beq
 [\delta\tau_{{\rm incoherent}}(\bth)]^2 =
 2\left(\frac{Gm}{c^3}\right)^2 \int_0^{D_l} (\delta y)^4 n dx
 \int_{y_{{\rm min}}}^{y_{{\rm max}}} \frac{2\pi y dy}{y^4} =
 \frac{2\pi}{5} \left(\frac{Gm}{c^3}\right)^2
 \frac{nD_l^5}{y_{{\rm min}}^2} \theta^4,
 \eeq
if one considers only the contribution to $(\delta\tau)^2$ from
foreground mass clumps that interact with the light rays during
redshifts $z < z_l$.  In arriving at the final expression of Eq. (9)
use was made of Eq. (7).  We may get rid of the dependence on
$y_{{\rm min}}$ by assuming that it equals the comoving distance at
which one or more clumps satisfy  the condition $y \leq y_{{\rm
min}}$, i.e.
 \beq
 y_{{\rm min}} = \frac{1}{\sqrt{\pi nD_l}}.
 \eeq
The contribution from background ($z_l < z < z_s$) clumps may
likewise be calculated and included.  One would then arrive at a
total variance of
\begin{equation}
c^2 (\delta\tau_{{\rm incoherent}})^2 = \frac{9}{160}
\left(\frac{H_0^4}{c^4}\right) \Omega_{{\rm cl}}^2 D_l^4 (D_l^2 +
D_{ls}^2) \theta^4,
\end{equation}
after employing the relation $Gnm = \sum_i Gn_i m_i =
3H_0^2\Omega_{{\rm cl}}/(8\pi)$, with $\Omega_{{\rm cl}}$ being the
mass density of clumps as a fraction of the critical density.

We intend to pursue an application of the above development, by
predicting the effect of external field galaxies on the time delay
between strong lensing images, for comparison with observations.
Before doing so, however, several remarks about
$(\delta\tau)^2_{{\rm random}}$ are in order.  Apart from the most
obvious fact that its final form scales only with one property of
the clumps, viz. $\Omega_{{\rm cl}}$, Eq. (11) is valid in the limit
$\delta y \ll y$. In terms of the comoving separation $d$ between
sources, Eq. (7), this implies (since $\delta y < d$ and $y_{{\rm
min}} < y$) that $d \ll y_{{\rm min}}$.  Now the images to be used
as testbeds have a typical comoving separation $d \lesssim$ 50 kpc,
i.e. a scenario under which the criterion is satisfied, because 50
kpc is not much larger than the size of a galaxy.  More elaborately,
if $y_{{\rm min}} \lesssim d$ one would `see' a galaxy when looking
at the sky along any direction: we assume this is not the case.

The next remark is that the sole role played by proximity clumps can
be seen from the dependence of $(\delta\tau)^2$ on $y_{{\rm min}}$
and not $y_{{\rm max}}$.  Thus, on the question of relative time
delay caused by galaxies one does not need to take account of large
(Mpc scale or more) distances over which mass compensation by galaxy
clustering is important. This justifies {\it a posteriori} our use
of a random clump ensemble without appealing to $P(k)$.  It also
provides the reason why the effect of voids on the time delay
fluctuations can be ignored: since the maximum separation between
the two rays is small compared with the typical inter-clump spacing
the void-to-void accumulation of the randomly varying second spatial
derivative of the void potential function is completely negligible
over such distance scales  (over much larger (CMB acoustic) distance
scales this phenomenon could play a role in lensing deflections via
mass clustering, see Holz \& Wald 1998, Seljak 1996).  The inclusion
of void effects will in principle {\it add} further signal to the
variance $\delta\tau_{{\rm incoherent}}$ for the presently assumed
(and justified) Poisson mass distribution, Thus even though the
neglected contribution is small, it does mean that our estimate of
$\delta\tau_{{\rm incoherent}}$ is conservative.

\section{Testing the dynamics of global geometry by strong lensing time delay}

The interpretation of cosmological time delay data is usually
confined to considerations of the delay within the strong lens
system and its immediate environs,
with the overall aim of inferring the Hubble constant $H_0$
from the observations (Refsdal 1964).  If e.g. the
lensing mass distribution is a singular isothermal sphere, the relative time
delay  between two images
at angular distances $\theta_A$ and $\theta_B$ from,
and on opposite sides of, the axis of symmetry is given by
\begin{equation}
\Delta\tau_{AB} = \tau_A - \tau_B =
\frac{1+z_l}{2} \frac{D_l D_s}{D_{ls}} (\theta_A^2 - \theta_B^2),
\end{equation}
In Eq. (12) it is assumed, of course, that A and B are images of the
same source, usually a time variable background quasar. The Hubble
constant clearly affects the delay via the distance dependence, viz.
$\Delta\tau_{AB} \sim D_l D_s/D_{ls} \sim H_0^{-1}$ (other
cosmological parameters also play a role because $D$ is a
multi-dimensional function, but their effects are minor, as noted by
Grogin \& Narayan 1996). Hence time delay measurements via light
curve alignment between images A and B of the quasar, coupled with
knowledge of redshifts, can in principle lead to a determination of
$H_0$.

To date approximately ten strong lensing systems with time delay
measurements are available (Saha et al 2006), in each case the delay
between two images separated by several arcseconds is typically
found to lie within the 10 -- 100 days range.  One pair of such
multiply lensed quasars with similar parameters,  SDSS J1004+4112
and HE0435-1223, were reported by Fohlmeister et al 2006 and
Kochanek et al 2006 respectively.  We shall employ this pair to
illustrate how the cosmological distribution of galaxies near the
light path can substantially enlarge the random uncertainty in the
value of $H_0$.  For SDSS J1004+4112 where the smallest observed
image separation, between images A and B, was $\theta \approx$ 4
arcsec and the comoving distances are $D_s =$ 4.77 Gpc, $D_l =$ 2.45
Gpc in an $\Omega_m =$ 0.3, $\Omega_{\Lambda} =$ 0.7 cosmology, Eq.
(11) gives $\delta\tau_{{\rm incoherent}} \approx$ 24 days if we
persist with the $\Lambda$CDM standard model by adopting its
breakdown of the matter budget to set $\Omega_{{\rm cl}} =$ 0.15
(specifically this assumes that half the baryons, hence
approximately the same fraction for dark matter also, of the low $z$
Universe resides in galaxies and their halos, see Fukugita 2004 and
Fukugita et al 1998).  In fact, taking the above parameters as
typical, one could proceed to recast Eq. (11) into a more convenient
form:
\begin{equation}
\delta\tau_{{\rm incoherent}} = 29.1 (h/0.7)^2 (\Omega_{{\rm
cl}}/0.15) (\theta/5~{\rm arcsec})^2 (D_l/2.5~{\rm Gpc})^3 \left[1 +
\left(\frac{D_{ls}}{D_l}\right)^2 \right]^{\frac{1}{2}}~~ {\rm
days}.
\end{equation}
Since the observed image delay of 38.4 $\pm$ 2.0 days is on par with
$\delta\tau_{{\rm random}}$, any estimation of $H_0$ that attributes
{\it all} the observed delay to physics within the strong lens
system would have caused this value to vary randomly from the truth
by almost 100 \%.  Other images of SDSS J1004+4112 can also be used
as testbeds: the separation of image B from C (also A from D) is
$\theta \approx$ 20 arcsec, while the expected time delay between
them is $>$ 560 days ($>$ 800 days for A and D), due solely to the
strong lens system itself.  From Eq. (13) we see that once again
$\delta\tau_{{\rm incoherent}}$ is comparable to these delays
because of its $\theta^2$ scaling.

A repetition of the above analysis to HE0435-1223, where $D_s =$
6.44 Gpc, $D_l =$ 1.74 Gpc and $\theta \approx$ 2 arcsec, results in
a similar though less drastic conclusion, viz. $\delta\tau_{{\rm
incoherent}} \approx$ 4.5 days versus the observed delay of 14.4
$\pm$ 0.8 days. The random error in $H_0$ here should then account
for an additional fluctuation $\delta H_0/H_0 \approx$ 30 \%.

In summary, the distribution, evolution, and mass budget of galaxies
as understood in the context of the standard cosmological model
leads to the prediction of a {\it random} (or
incoherent) relative delay between
the light  arrival times from two images of
a strongly lensed background quasar
comparable with the observed delay.  Since
the latter has routinely been interpreted as an effect caused principally
by the gravitational field of the lens, and moreover a value of
$H_0$ is derivable from it if perturbations
outside the strong lens are absent, the question of whether additional
and hitherto unknown dynamics are responsible for the global flatness
of space could be addressed by examining the statistical variation in
the $H_0$ values that emerge
from a large number of strong lensing
delay measurements, when such a database becomes available.

If this variation distributes around a value of $H_0$ that agrees
with other methods of determination, with a standard deviation
matching the expectation from Eq. (10), it would imply that for the
first time the {\it ensemble} gravitational effect of many galaxies
spread over cosmological distances has been detected.  If, on the
other hand, the variation distributes tightly around the accepted
value of $H_0$ with no room for extra perturbations, the possibility
of a new physical phenomenon that complements (even replaces) the
law of gravity as the distance scale becomes large must then be
inevitable.  The fact that there has not been any report of a
significant deviation from the $h =$ 0.7 mark in the value of $H_0$
as determined by this type of time delay technique may already be
pointing to the absence of the random effect discussed here.  If
this turns out to be really the case, the stability of the global
geometry as revealed here would present a serious challenge to
cosmology.

We end by pointing out that the possible influence of foreground
matter on the measurement of $H_0$ was considered in a recent work
(Fassnacht et al 2006) under the scenario of this matter being
clumped into several foreground groups of galaxies.  The present
paper, however, calculates for the first time the effect of a random
ensemble of many foreground clumps  on strong lensing time delay; we
then demonstrated that this introduces a large scatter in the
resulting value of $H_0$.  The cause of such a scatter stems mainly
from light skirting clumps without passing through them, i.e.
cosmological time delay data contain precious information on {\it
shear}.

\acknowledgements
The author thanks an anonymous referee for very helpful criticisms
towards improving this paper.

\end{document}